1

# Mesoscopic Quantum Emitters from Deterministic Aggregates of Conjugated Polymers

*Thomas Stangl[1], Philipp Wilhelm[1], Klaas Remmerssen[2], Sigurd Höger[2], Jan Vogelsang[1,*] and John M. Lupton[1]*

[1]Institut für Experimentelle und Angewandte Physik, Universität Regensburg, Universitätsstrasse 31, 93053 Regensburg, Germany

[2]Kekulé-Institut für Organische Chemie und Biochemie, Universität Bonn, Gerhard-Domagk-Strasse 1, 53121 Bonn, Germany

* Corresponding author: Jan Vogelsang, Institut für Experimentelle und Angewandte Physik, Universität Regensburg, Universitätsstrasse 31, 93053 Regensburg, Germany, E-Mail: jan.vogelsang@physik.uni-regensburg.de

**ABSTRACT**

An appealing definition of the term "molecule" arises from consideration of the nature of fluorescence, with discrete molecular entities emitting a stream of single photons. We address the question of how large a molecular object may become by growing deterministic aggregates from single conjugated polymer chains. Even particles containing dozens of individual chains still behave as single quantum emitters due to efficient excitation energy transfer, while the brightness is raised due to the increased absorption cross-section of the suprastructure. Excitation energy can delocalize between individual polymer chromophores in these aggregates by both coherent and incoherent coupling, which are differentiated by their distinct spectroscopic fingerprints. Coherent coupling is identified by a ten-fold increase in excited-state lifetime and a corresponding spectral red shift. Exciton quenching due to incoherent FRET becomes more significant as aggregate size increases, resulting in single-aggregate emission characterized by strong blinking. This mesoscale approach allows us to identify intermolecular interactions which do not exist in isolated chains and are inaccessible in bulk films where they are present but masked by disorder.

**SIGNIFICANCE STATEMENT**

Bright and stable single-photon sources, based on molecular objects, have contributed to exploring the foundations of quantum mechanics and measurement theory. Since the photon emission rate scales with molecular size, the most direct approach to increasing brightness is to enlarge the molecular object itself. But how big can it become while still retaining molecular characteristics of a deterministic photon source? We tackle this question by growing defined aggregates out of single chains of a conjugated polymer. Multichain aggregates show discrete single-photon emission, even when the individual chains display multi-photon emission. Single-aggregate spectroscopy reveals how coherent and incoherent excitonic intermolecular coupling mechanisms, known from the bulk, evolve with aggregate size.

\body

**Introduction**

A molecule, as the smallest entity of a material, is a deterministic discrete object. A simple optical technique can be devised to test this discreteness of molecules: photon antibunching, the deterministic fluorescence emission of a stream of individual photons, one at a time (1-3). By dissolving a molecular material and diluting it to ever smaller concentrations, recording the fluorescence with a microscope objective, and passing the light through a beam splitter onto two different photodetectors, photon coincidence rates on the two detectors are measured. Since a single photon cannot be at two places at once, discrete emission of single photons is easily observed in a drop of the coincidence rate at zero delay time between the two detector channels. This test of molecular discreteness begs a simple question: how large can a molecular object become for it to still behave as a perfect quantum emitter? Recently, antibunching has been demonstrated from large π-conjugated macrocycles (4) over 6 nm in diameter, and from comparably-sized natural light-harvesting complexes (5). Less pronounced antibunching has also been observed from some multichromophoric conjugated polymers of comparable molecular weight (6). Since the ease of deterministic synthesis of ultra-large π-conjugated complexes deteriorates rapidly with size, one may consider instead growing molecule-like objects by van-der-Waals bonding to small aggregates – the "*molecular mesoscopic*" approach. Such aggregates can be grown in a controlled way by single-molecule solvent vapor annealing (7), raising the question of what the fundamental size scale is for which a transition from molecular to bulk behavior occurs.

Fluorescent molecules in an aggregate can interact by coherent (8, 9) or by incoherent excitation energy transfer (EET) (10). Both processes can lead to a change in fluorescence lifetime and spectrum and are therefore hard to distinguish in ensemble measurements. On the single-chain to single-aggregate level, differentiation is much easier: coherent interchromophoric coupling between parallel chromophores leads to a delocalization of excitation energy, resulting, in the simplest case, in an H-aggregate-like spectral shift: the excited-state energy level splits in two, with a redistribution of oscillator strength to the higher-lying state (11). In a molecule in the solid state, with inhibited motion and diffusion, emission becomes excimer-like, broadening and shifting to the red (12). Since oscillator strength is lost from the lowest-energy transition, an increase in fluorescence lifetime is observed – provided, however, that there is no incoherent EET (i.e. FRET) to molecular quenching sites which induce non-radiative decay (13, 14).

We recently approached the investigation of intermolecular interactions on the sub-ensemble level by designing model systems with parallel chromophores within a single molecule (13). Excimer-like emission evolves for parallel oligomers spaced fewer than 5 Å apart (14). However, long-range interactions over mesoscopic distances, such as incoherent EET which can persist over tens of nanometers (15), or the coherent coupling of multiple chromophoric units at once, remain inaccessible in these small model systems. We therefore aim to isolate highly ordered interchain aggregates to study and compare their electronic properties with those of single chains by employing single-aggregate and single-molecule spectroscopy. Fig. 1 illustrates the basic approach pursued to building molecular aggregates from the bottom up. We use poly(*para*-phenylene-ethynylene-butadiynylene) (PPEB), since this material-class is well known to give rise to excimer-like emission in the solid state (16) as seen in a strong spectral shift to the red from solution phase to solid film. In a bulk film, both ordered (red) and disordered (green) domains exist. Dissolving the bulk to the level of single molecules, i.e. single polymer chains, gives rise to spatially discrete objects which can contain multiple chromophores. The challenge lies in obtaining small, morphologically well-defined and spatially isolated aggregates of a particular size: this is the "*molecular mesoscopic*" dimension between the individual molecule and the bulk film. This challenge can be met by *in situ* solvent vapor annealing (7) (SVA) to form ordered multi-chain aggregates of predetermined average size. Remarkably, even large aggregates can retain the quantum-optical characteristics of a single molecule, while showing clear signatures of coherent interchromophoric coupling (H-aggregate formation) and incoherent interchromophoric EET.

Before discussing the results, it is crucial to note that the spectral shift and photoluminescence (PL) lifetime changes observed on the single-molecule level only provide a metric for establishing that coherent coupling occurs but do not allow an extraction of the actual spatial coherence size of the molecular exciton. Although theoretical models have been proposed to extract the coherence size from the change in spectral properties (17), such models assume perfect order, i.e. intermolecular arrangement and spacing, of the individual chromophores. It is unlikely to achieve such order with large π-conjugated chromophores as the "monomer" building blocks of the aggregate. Our recent studies of H-aggregate excimer-like emission from single model dimer structures revealed that the spectroscopic observables can fluctuate with time (14), implying changes in molecular equilibrium conformations. We therefore stress that, while the change of spectral shape, red-shift and increase of lifetime do provide unambiguous evidence for coupling between at least two chromophores, the spectral characteristics cannot be used to quantify spatial coherence.

**Results and Discussion**

**Building ordered aggregates by SVA.** The average size and morphological order of aggregates can be controlled by the solvent used for SVA (7). PPEB molecules with a number-average molecular weight of $M_n = 40$ kDa and a polydispersity index of PDI = 1.46 were dissolved in a poly(methyl-methacrylate) (PMMA) / toluene solution and diluted to a concentration of ~$10^{-11}$ M, above typical concentrations for single-molecule experiments (14). The solution was spin-coated to obtain a 200-250 nm thick PMMA film with single molecules uniformly distributed. During SVA, the PPEB/PMMA film is in a heterogeneous mixture of solid- and liquid-like phases in which single chains undergo diffusion, as illustrated in Fig. 1. Diffusion of single chains leads to aggregation. After 30 minutes of SVA, the sample was dried with nitrogen, immobilizing molecules and aggregates.

The films were scanned by a confocal microscope to identify the positions of single molecules and aggregates. This information was used to localize one single molecule, or aggregate, at a time in the diffraction-limited excitation spot at 405 nm. A dichroic mirror split the PL into two detection channels above and below 532 nm, as shown in Fig. 2A, to separate emission from aggregates (orange-red) and isolated chains (blue-green) remaining in the film. Fig. 2B-E shows a series of confocal images for samples with different solvent mixtures used for SVA. Green denotes PL detected below 532 nm, originating from isolated chains, and red the aggregate PL above 532 nm. Fig. 2B shows an image of 20×20 µm² area of single chains in PMMA at single-molecule concentration ($10^{-12}$ M). Exclusively green emission from isolated chains is seen in diffraction-limited spots. Aggregates can be formed by selectively dissolving the PMMA matrix in acetone, in which the conjugated polymer is insoluble. Addition of a suitable polymer solvent (chloroform) to acetone promotes solubility in the swollen PMMA matrix, driving the formation of larger aggregates. Panels C-E show corresponding images of samples after 30 minutes of SVA with varying acetone to chloroform vapor ratios denoted in brackets (see Methods for details on sample preparation). A second population of orange to red diffraction-limited spots evolves after SVA with 100 % acetone vapor (panel C), indicating that these spots emit significantly above 532 nm. After SVA with a vapor of 90:10 acetone:chloroform ratio (panel D), the overall number of spots decreases while the brightness of the red spots increases. This effect is more pronounced after SVA with a 80:20 ratio (panel E) (7). The increasing aggregate size with increasing solvent-polymer miscibility can be rationalized in terms of Ostwald ripening (18), which describes the particle size up to which stable aggregates can be formed. All images are shown on the same intensity scale, which leads to a blurring of the brightest spots for the largest aggregates. In all cases, the individual spots are diffraction limited.

The fraction of red PL, $F_{red}$, is calculated as $I_{red}/(I_{red} + I_{green})$ and shown for each particle in the histograms. Prior to SVA, only $F_{red} < 0.3$ values are found, whereas a significant number of particles exhibit values above 0.3 after SVA. Based on this observation, we count the number of single chains (green) and aggregates (red) per image before and after SVA, providing the average number of chains in an aggregate as stated in each image. Using this preparation method for isolated aggregates of different sizes, we can now compare the spectroscopic properties of single chains and aggregates.

**Spectroscopic properties of individual polymer chains *vs*. single aggregates.** First, we compare the morphology of isolated chains and aggregates using excitation polarization spectroscopy, which reports on the overall anisotropy in absorption (19). The excitation beam is linearly polarized in the sample plane and the polarization is rotated while recording the PL intensity as depicted in Fig. 3A. The excitation polarization modulation depth, $M$, is obtained by fitting the PL intensity, $I$, as a function of polarization angle, $\theta$, to Malus' law,

$$I(\theta) \propto 1 + M\cos 2(\theta - \Phi)$$

where $\Phi$ is the orientation angle of the molecular transition dipole moment for maximal PL. For each single spot, $M$ was acquired, yielding a histogram as shown in Fig. 3B. Since molecular weight affects chain morphology (20-22) we compared two different weights of PPEB. The histogram shaded in dark green shows the distribution of modulation depth values for short chains ($M_n \approx 40$ kDa). The light shaded histogram reports on long chains with $M_n \approx 210$ kDa. The anisotropy decreases as chain length increases since longer chains fold more (20). This behavior is in contrast to that of isolated aggregates containing, on average, 12 chains, shown in Fig. 3C. Only spots with PL emission above 532 nm were taken into account to separate the aggregates from remaining single chains. The resulting modulation depth histogram has a maximum at $M\sim0.8$. We conclude that PPEB undergoes aggregation-induced ordering during SVA, leading to the first building blocks of crystalline structures which characterize the bulk film (23, 24).

The emergence of well-ordered aggregates is further supported by comparing typical PL spectra and transient PL decays of an individual aggregate and a single chain in Fig. 3D, E. Whereas the PL spectrum of single chains is well structured with a 0-0 peak at 465 nm and a vibronic progression reaching up to 520 nm (Fig. 3D, green spectrum), the spectrum of the aggregate is less structured with a suppressed 0-0 transition around 500 nm and a vibronic progression extending to 700 nm (Fig. 3D, red spectrum). Simultaneously, the PL lifetime is increased ten-fold from 0.5 ns to 5.3 ns in going from single chains to aggregates. These observations are explained within the framework of excimer-like luminescence from H-aggregates, i.e. coherent interchromophoric coupling (12, 14, 25-27). Coupling of adjacent chromophores oriented in parallel to one another in the excited state leads to an energetic splitting of the excited state, where the transition dipole moment vanishes in the lower-lying energy level as sketched in the inset of Fig. 3D. The PL therefore shifts to the red concomitant with a decrease in radiative rate, leading to an increase in PL lifetime provided that the fluorescence quantum yield does not change (11, 28).

**Role of coherent and incoherent interchromophoric coupling in different aggregate sizes.** To substantiate the correlation between red-shifted PL and increased PL lifetime, we recorded the $F_{red}$ value and the PL lifetime, $\tau_{PL}$, for each spot in the microscope image. Fig. 4 shows scatter plots between $F_{red}$ and $\tau_{PL}$ for the different samples: for 326 isolated chains in panel A, $F_{red}$ is narrowly distributed between 0.07-0.25 with 0.4 ns < $\tau_{PL}$ < 1.2 ns. This distribution changes dramatically upon SVA. For the smallest aggregates with ~12 chains, the $F_{red}$ values for 583 aggregates scatter between 0.1 and 0.85 (panel B), with a strong correlation between $F_{red}$ and $\tau_{PL}$, which can be as large as 6.3 ns. Annealing with a solvent mixture (90:10 acetone:chloroform) leads to an average size of 18 chains per aggregate. For these particles, the scatter plot of 529 spots (Fig. 4C) indicates a bimodal distribution. The first distribution shows $F_{red}$~0.2 and small $\tau_{PL}$ values from 0.4-2 ns, similar to the isolated chain (the "monomer"). The second distribution is characterized by $F_{red}$~0.8 with $\tau_{PL}$ values scattering strongly between 0.8 ns and 6 ns. Even larger aggregates with, on average, 54 chains can be formed by annealing with an 80:20 solvent ratio; the corresponding scatter plot of 299 particles (Fig. 4D) shows $F_{red}$ values grouped almost exclusively around 0.8, again accompanied by a wide range of $\tau_{PL}$ values between 1-5 ns.

We conclude that the sample with the smallest aggregate size of, on average, 12 chains consists of isolated chains (green spots in Fig. 4B), small or loosely bound aggregates (yellow spots), and large or strongly bound aggregates (red spots). The strong correlation between PL red shift and increased PL lifetime implies the emergence of a coherently coupled interchromophoric excited state within polymer aggregates. Small or loosely bound aggregates vanish in samples with increasing aggregate size. However, the scatter of PL lifetimes increases with increasing aggregate size. This effect can be explained by interchain EET and luminescence quenching (10). In an aggregate, the probability of generating a fluorescence quencher such as a hole polaron (10) is greater than in an isolated chain since more molecular units are involved in absorption and longer-range charge transfer can occur, thus raising the susceptibility to exciton quenching (29, 30).

To investigate the PL quenching mechanism in the aggregates, we examined the correlation of PL intensity and lifetime for the largest aggregates (54 chains average). To ensure complete aggregation of the chains, only spots with $F_{red} > 0.7$ were selected, marked in blue in Fig. 4D. The corresponding scatter plot is shown in Fig. 5A. Short PL lifetimes correspond to low PL intensities. In contrast, long lifetimes arise for both high and low PL intensities, corresponding to unquenched and quenched aggregate PL. Since there is an inherent distribution in aggregate size, a direct correlation between PL lifetime and intensity is masked in the statistical analysis of many single aggregates. However, this averaging is overcome by considering the temporal dynamics in PL lifetime and intensity of a single aggregate, as demonstrated in Fig. 5B: a reduction in PL intensity correlates directly with a drop in PL lifetime, implying a decrease in quantum yield due to increased non-radiative rate. A large aggregate from the sample containing, on average, 54 chains was placed in the confocal excitation area, and the PL intensity, lifetime and $F_{red}$ values were recorded simultaneously.

The PL intensity shows strong fluctuations between discrete intensity levels over timescales of seconds. Quenching events are as strong as 80 % of the maximum PL intensity, and are correlated with PL lifetime, which fluctuates between ~4.1 ns (at maximum intensity) and ~1 ns (at minimum intensity). At the same time, $F_{red}$ remains constant at ~0.8, implying that the spectrum and thus the coherent interchromophoric coupling does not change during dynamic PL quenching events (30-33).

The correlation between PL intensity and lifetime with increasing aggregate size implies that the fast PL decay generally seen in bulk films of PPE-based materials (34), where coherent interchromophoric coupling induces a red shift, arises from photochemical quenchers (34, 35), which have a strong effect on fluorescence over a large area surrounding the quencher. The strong blinking observed here in multi-chain aggregates implies long-range interchain EET (7), which should also result in efficient singlet-singlet annihilation with subsequent single-photon emission (36). Such behavior was previously reported for highly ordered *single* P3HT chains in conjunction with efficient singlet-triplet annihilation (21), *single* cyano-substituted polyphenylene vinylene chains (37), and synthetic and natural multi-chromophoric light-harvesting systems (5, 36, 38-41), but has not been observed for large *multi*-chain aggregates. Interchain EET can be resolved through the photon statistics in fluorescence and quantified by the degree of photon antibunching relative to isolated chains.

**Single photon emission from multi-chain aggregates.** A single chain can generally be described by a series of more-or-less strongly interacting chromophores (42, 43). Upon excitation with light, multiple chromophores can enter the excited state at once. If the chromophores are independent of each other—that is, there is no dipolar coupling or electron

tunneling—multiple photons are emitted simultaneously. However, even in an unfolded chain, energy transfer between chromophores will usually occur so that one chromophore can transfer its energy to an excited state of another, even if the latter is already excited. Since double excitation of a molecular unit changes configuration coordinates, subsequent excitation above the fundamental gap dissipates energy non-radiatively. Such singlet-singlet annihilation, driven by energy transfer, leads to photon antibunching from multichromophoric systems (36, 38). The quality of photon antibunching as a function of molecular size is therefore directly related to energy transfer within the multichromophoric aggregate (21).

To reveal interchromophoric interactions, we measured the statistics of fluorescence photons by splitting the detection path onto two detectors which yields the number of correlation events, $N$, in dependence of the difference in photon arrival times, $\Delta\tau$, between the two detectors, as sketched in the inset of Fig. 6B. Fig. 6A shows a typical PL transient of a short PPEB chain (40 kDa sample) with strong blinking accompanied by a gradually decaying PL intensity. Panel B plots the correlation events acquired with laser pulses separated by 50 ns. The ratio of the magnitude of the central peak at $\Delta\tau = 0$ to that of the lateral peaks, $N_C/N_L$, provides a measure for the degree of photon antibunching (depicted as blue dashed lines in the center panels). A ratio of $N_C/N_L = 1$ implies an infinite number of independently emitting chromophores in the excitation spot, whereas $N_C/N_L = 0$ corresponds to a single effective chromophore (44); in a multichromophoric aggregate such an observation translates to near unity EET efficiency within the particle (36, 38). For the example shown in Fig. 6A, a ratio of $N_C/N_L = 0.44$ is determined, which approximately corresponds to two independently emitting chromophores averaged over the entire acquisition time of 15 s (44). The $N_C/N_L$ ratio for 160 single chains is plotted in panel C. The distribution is broad with a maximum around 0.3. The scatter most likely reflects the molecular weight distribution. A higher $N_C/N_L$ is found for longer PPEB chains (210 kDa sample), implying more active chromophores. An intensity trace for such a molecule (Fig. 6D) leads to a ratio of $N_C/N_L = 0.71$ (panel E), with the distribution for 203 chains showing a maximum around 0.7 (panel F).

PPEB aggregates yield more surprising results. Fig. 6G shows a typical PL transient of a PPEB aggregate (~12 40 kDa polymers/aggregate), with the corresponding cross-correlation shown in panel H. Strong photon antibunching is found with a $N_C/N_L$ ratio of 0.16; the object therefore closely resembles a single-photon source, although it consists of multiple chains, which by themselves do not show strong photon antibunching. The distribution of $N_C/N_L$ values between particles (panel I) retains the breadth seen in isolated chains (panel C), but the absolute values are significantly reduced, with a maximum around 0.1, implying virtually perfect photon antibunching from single-chain aggregates.

**Conclusions**

Based on these observations, we draw the following conclusions: (i) slow aggregation by SVA leads to highly ordered aggregates in which coherent coupling between single chains evolves (Fig. 2 and 3). (ii) This coupling can best be described in the context of the formation of an excited state involving multiple chromophores with excimer-like emission of substantial oscillator strength, and leads to a strong red shift in PL and a decrease in radiative rate (Fig. 3 and 4). (iii) The coherent coupling between at least two chromophores along with the high degree of structural ordering in the multi-chain aggregates promotes effective EET, which does not occur at the single-molecule level (Fig. 4 and 5). (iv) EET is so effective that tens of chains couple together to behave as a single quantum emitter (Fig. 6). (v) The formation of quenchers becomes more likely with increasing aggregate size, opening up additional non-radiative decay channels observed by a reduction in the PL lifetime (Fig. 5). This effect is the likely reason why the long PL lifetime, reported here for single aggregates, is not observable in bulk PPE-based films (34) even though the emission spectra are very similar. Single-aggregate spectroscopy of conjugated polymers can therefore bridge the gap between isolated chains and bulk films, revealing mesoscopic interactions which are not apparent in both extreme states of the material. Unexpected phenomena such as deterministic single-photon emission evolve in this mesoscopic size regime provided chain ordering is well controlled. The strong spectroscopic differences between single chains and aggregates provide a unique observable for studying nucleation and crystallization pathways of conjugated polymers *in situ*, opening new experimental routes to polymer physics in general. Finally, we stress that our approach to controlling morphology of single emitters *in situ* is applicable to any form of emitter, be it a colloidal quantum dot or a phosphorescent molecule. The recent interest in the surprising morphology and counter-intuitive orientational anisotropy of triplet emitters in OLEDs (45), which controls light out-coupling efficiency, will provide a rich environment for applying the techniques presented here.

**Methods**

**Sample fabrication.** Poly(*para*-phenylene-ethynylene-butadiynylene) (PPEB) was synthesized as described in detail elsewhere (20), and purified using a gel-permeation chromatograph (GPC) to obtain two samples with a number average molecular weight $M_n = 40$ kDa with a PDI of 1.46 and $M_n = 210$ kDa with a PDI of 1.47. Poly(methyl-methacrylate) (PMMA, $M_n = 46$ kDa, PDI = 2.2) was purchased from Sigma-Aldrich. Isolated chains of PPEB molecules were embedded in a PMMA host matrix by dynamically spin-coating from toluene on glass cover slips, which were cleaned according to a published procedure (4). The PMMA film thickness was 200-250 nm, and the concentration of PPEB in solution before spin-coating was ~$10^{-12}$ mol·l$^{-1}$ and ~$10^{-11}$ mol·l$^{-1}$ for the single-molecule and aggregate samples, respectively. The samples were incorporated into a gas flow cell and annealed under solvent vapor with different acetone to chloroform ratios for 30 min to prepare differently sized aggregates. Details of the SVA process for aggregation can be found in ref. (7).

**Scanning confocal microscope.** The samples were investigated with a scanning confocal microscope based on an Olympus IX71 (14). Excitation was carried out by a fiber-coupled diode laser (PicoQuant, LDH-D-C-405) at 405 nm under pulsed excitation with a repetition rate of 20 MHz for photon statistics measurements or 40 MHz for PL lifetime measurements. The excitation light was passed through a clean-up filter (AHF Analysentechnik, HC Laser Clean-up MaxDiode 405/10) expanded and collimated via a lens system to a beam diameter of ~1 cm and coupled into an oil-immersion objective (Olympus, UPLSAPO 60XO, NA = 1.35) through the back port of the microscope and a dichroic mirror (AHF Analysentechnik, RDC 405 nt) for confocal excitation with an intensity set to 50 W/cm². Fluorescence images (size of 20×20 µm², integration time 2ms/pixel with a resolution of 50 nm/pixel) were recorded by stage scanning (Physik Instrumente, model P-527.3CL). The fluorescence signal passed a 50 µm pinhole and fluorescence filter (AHF Analysentechnik, Edge Basic LP 405 long pass filter) and was split by a dichroic mirror (AHF Analysentechnik, z532rdc) and detected by two avalanche photodiodes (APDs, PicoQuant, τ-SPAD-20) connected to a time-correlated single-photon counting module (TCSPC, PicoQuant GmbH. HydraHarp 400) for separating single chain and aggregate emission (Fig. 2). The images were evaluated by a home-written LabView software capable of automatically detecting single spots for which the fraction of red emission, $F_{red}$, was calculated and simultaneously the PL lifetime was extracted (Fig. 4). Alternatively, the fluorescence signal was split by 70/30 beam splitter to simultaneously detect 30% of the PL on an avalanche photodiode (Micro Photon Devices S.r.l., PDM Series) connected to the TCSPC unit and 70% on a spectrograph (Andor technology plc., SR-303i-B) coupled with a CCD camera (Andor Technology plc., DU401A-BV) to obtain PL decays and spectra (Fig. 3D, E) from spots which were subsequently placed inside the excitation focus. For photon statistics measurements the fluorescence signal was split by a 50/50 beam splitter and detected by two avalanche photodiodes both connected to the TCSPC unit to record time-tagged photon arrival times, which were further analyzed by a home-written LabView program.

**Excitation polarization spectroscopy.** The same microscope was used in wide-field excitation mode for the excitation polarization measurements, which are shown in Fig. 3A-C. Details can be found in ref. (14). The fluorescence signal passed through an additional filter (AHF Analysentechnik, Edge Basic LP 532 long pass filter) to select only aggregates emitting above 532 nm for the histogram shown in Fig. 3C.

**Acknowledgments**

The authors are indebted to the European Research Council for funding through the Starting Grant MolMesON (305020), to the Volkswagen Foundation for continued support of the collaboration and thank the German Science Foundation for support through the SFB 813.

**Author contributions**

T.S., P.W. and J.V. conceived, designed and performed experiments and analyzed the data. K.R. and S.H. designed and synthesized the compounds. J.V. and J.M.L. wrote the manuscript.

Correspondence and requests for materials should be addressed to:

Dr. Jan Vogelsang; e-mail: jan.vogelsang@physik.uni-regensburg.de

**Competing financial interests**

The authors declare no competing financial interests.

**Figure captions**

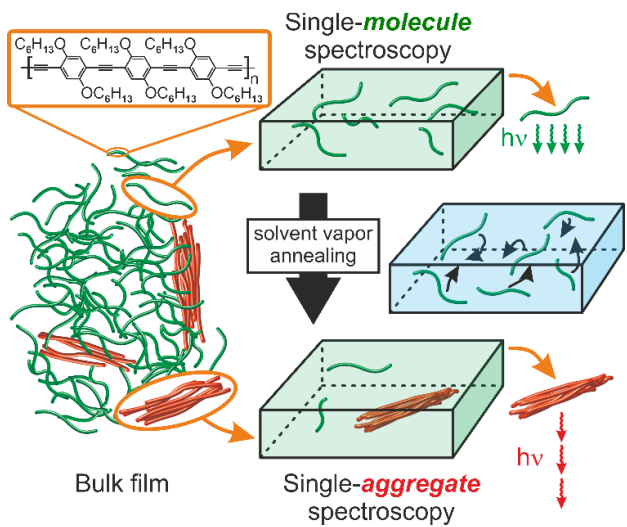

**Fig. 1.** Single-molecule and single-aggregate spectroscopy by solvent vapor annealing. A schematic drawing of a poly(para-phenylene-ethynylene-butadiynylene) (PPEB) conjugated polymer bulk film is depicted on the left (chemical structure shown on top) with amorphous (green) and aggregated or crystalline regions (red). For single-molecule spectroscopy, single conjugated polymer chains (green) are isolated by embedding highly dilute concentrations of conjugated polymer in a non-fluorescent polymer host matrix (shaded light green, top). Single isolated aggregates are formed during solvent vapor annealing (SVA) with a suitable solvent (shaded light blue, middle), which allows for the diffusion of single chains inside the swollen

host matrix and facilitates aggregation. Isolated and well-ordered aggregates (red) are formed inside the host matrix following the SVA process. One of the main results is depicted on the right: a single conjugated polymer chain emits multiple photons simultaneously, whereas an aggregate, consisting of tens of chains, radiates one photon at a time.

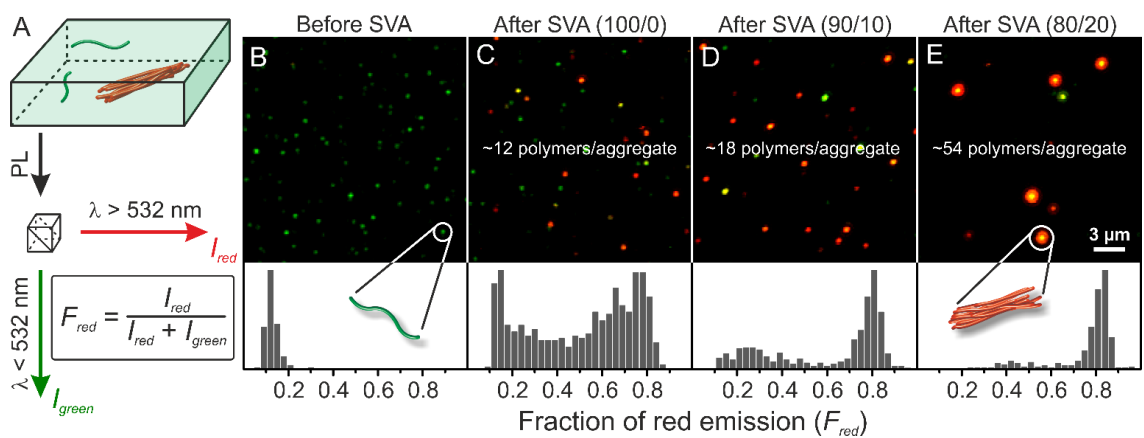

**Fig. 2.** Controlled growth of conjugated polymer aggregates from isolated single chains under different processing conditions. (A) Schematic illustrating the splitting of single-spot PL into two detection channels, for photons with $\lambda > 532$ nm (denoted $I_{red}$) and $\lambda < 532$ nm (denoted $I_{green}$). The fraction of red emission for each single spot is defined as $F_{red}$. The acetone to chloroform ratio used for SVA of single chains is given in parentheses above each confocal scanning microscope image (B-E). The corresponding $F_{red}$ values are shown in histograms below each image alongside the average number of polymer chains per aggregate.

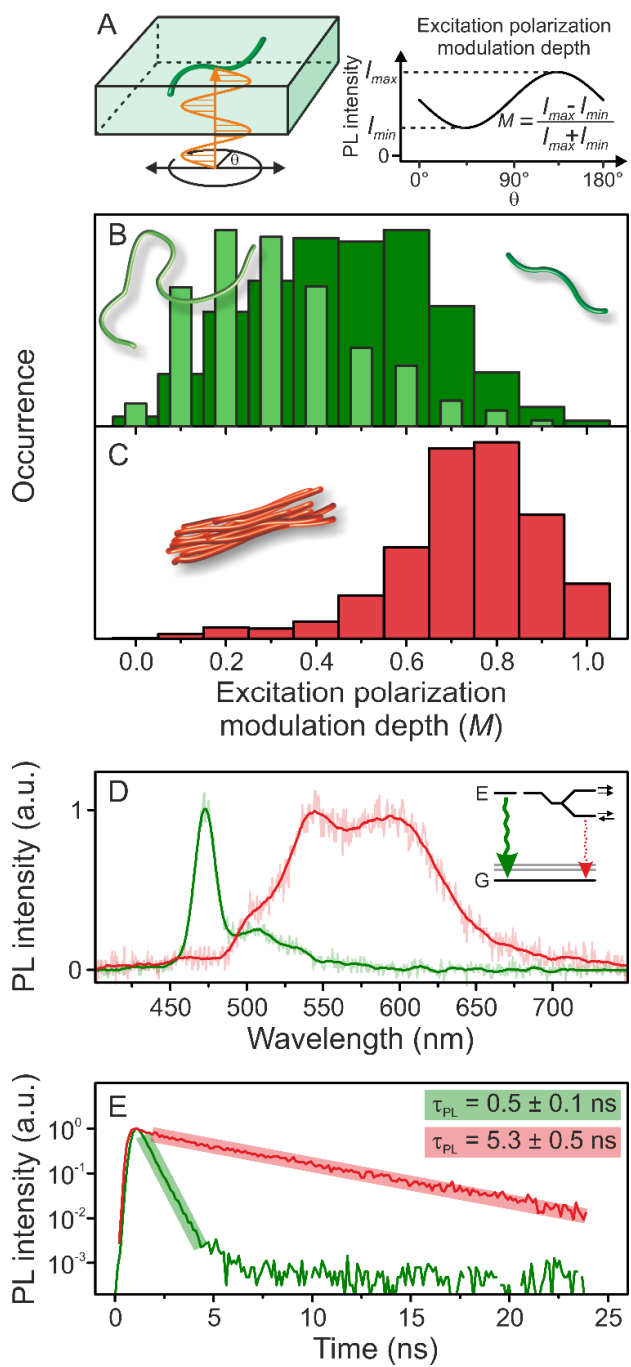

**Fig. 3.** Structural and spectroscopic properties of single polymer chains and isolated aggregates. (A) Schematic representation of the excitation polarization measurement procedure and definition of modulation depth, M. (B) M values for single chains with an average molecular weight of 40 kDa (dark green, 562 spots measured) and 210 kDa (light green, 1686 spots measured). (C) M values for single aggregates ($\lambda_{PL}$>532 nm, 1340 spots

measured). (D, E) Normalized PL spectra and transient PL decay of a single chain (green) and aggregate (red). Spectrum and decay were recorded simultaneously. The inset in (D) depicts the energetic splitting of the excited state, E, due to coherent coupling of neighboring chromophores in an aggregate. The green and red arrows represent the excited to ground state transitions within an isolated chain and an aggregate, respectively. The aggregate transition is red-shifted and retarded since the transition dipoles of the lower-energy state cancel out.

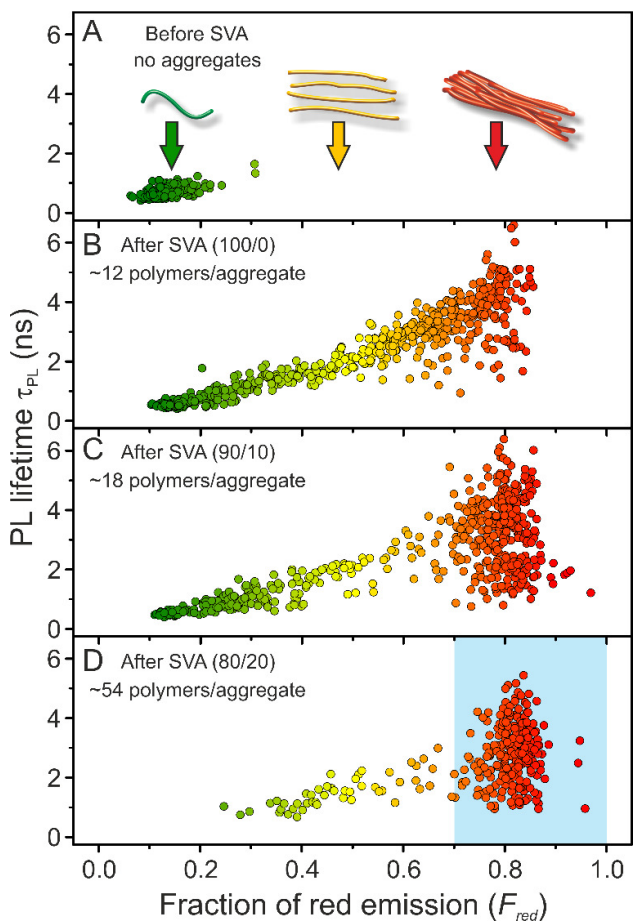

**Fig. 4.** Evolution of spectral characteristics from single chains to deterministic aggregates of increasing size. (A) Scatter plot of PL lifetime, $\tau_{PL}$, and fraction of red emission, $F_{red}$, for single chains and (B-D) isolated PPEB aggregates formed by SVA with different acetone to chloroform vapor ratios, denoted in brackets. The average number of chains per aggregate is stated in each panel. The colors scale from green (isolated chain) over yellow/orange (weakly

aggregated chains) to red (ordered aggregates). The red-most spots in the light blue shaded area in panel D were used for the further analysis in Fig. 5A.

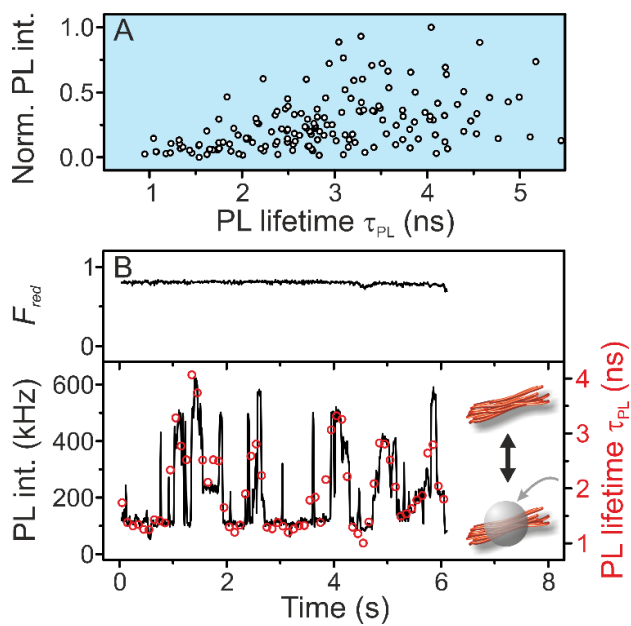

**Fig. 5.** Quenching of PL in large aggregates. (A) Scatter plot of PL intensity and lifetime for large multi-chain aggregates (average of 54 chains), exhibiting a fraction of red emission of $F_{red} > 0.7$. These particles correspond to the data spots shaded light blue in Fig. 4D. (B) Temporal fluctuations in PL intensity and lifetime of a single aggregate (time binning 10 ms, PL lifetime values derived from a running average of 500 ms stepped in 5 ms increments). The emission spectrum of the aggregate does not change during intensity and lifetime fluctuations as seen in the corresponding $F_{red}$ values.

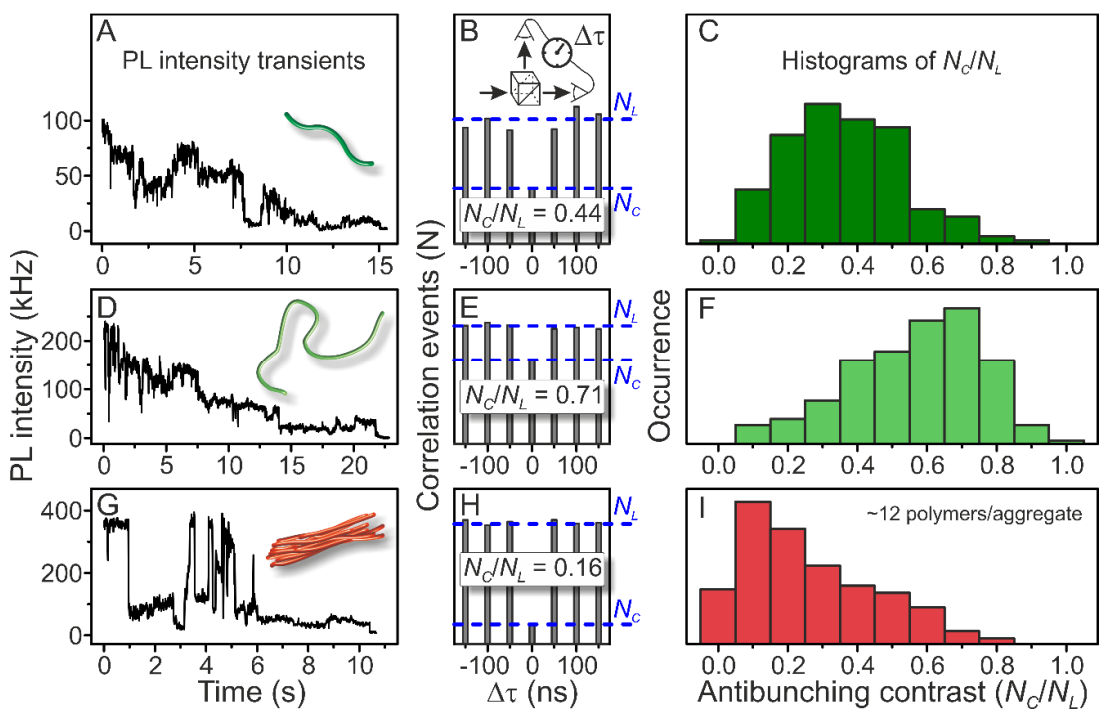

**Fig. 6.** Photon antibunching from single multi-chain aggregates due to efficient interchain energy transfer and singlet-singlet annihilation. Typical PL intensity transients (blinking) are shown on the left for short (~40 kDa, (A)) and long (~210 kDa, (D)) single PPEB chains, and aggregates consisting, on average, of 12 polymers (G). (B, E, H) The corresponding photon statistics in emission are shown in terms of the correlation events, $N$, of two photodetectors in the emission pathway. The molecules were excited by laser pulses (20 MHz repetition rate), allowing for the difference $\Delta\tau$ in photon arrival times between the two detectors to be controlled. The ratio between the signal of the center peak, $N_C$, and lateral peaks, $N_L$, is stated in each panel, which specifies the contrast of photon antibunching. (C, F, I,) Histograms of $N_C/N_L$ values for the three samples measured over 160, 203 and 176 spots, respectively.